\documentclass[modern]{aastex63}

\newcommand\ha{{H$\alpha$}}
\newcommand\hb{{H$\beta$}}
\newcommand\hst{{$HST$}}
\newcommand\spitz{{$Spitzer$}}

\newcommand\kms{\:\rm{km\,s^{-1}}}

\newcommand\FLUX{\:{\rm erg\:cm^{-2}\:s^{-1}}}

\newcommand\OiiiL{[\ion{O}{3}] $\lambda\lambda 4959,5007$}
\newcommand\SiiL{[\ion{S}{2}] $\lambda\lambda 6717, 6731$}
\newcommand\NiiL{[\ion{N}{2}] $\lambda\lambda 6548, 6583$}

\newcommand\FeiiL{[\ion{Fe}{2}] $\lambda 1.644 ~ \mu$m}

\newcommand\sii{[\ion{S}{2}]}
\newcommand\nii{[\ion{N}{2}]}
\newcommand\oi{[\ion{O}{1}]}

\newcommand\oiii{[\ion{O}{3}]}
\newcommand\feii{[\ion{Fe}{2}]}

\newcommand{\MSOL}{\mbox{$\:M_{\sun}$}}  

\received{July 7, 2020}
\revised{AUgust 10, 2020}

\shorttitle{HST Observations of N103B}
\shortauthors{Blair et al.}

\usepackage{natbib}
\usepackage{comment}
\usepackage{todonotes}
\usepackage{graphicx}

\begin{document}

\title{Imagery and UV Spectroscopy of the LMC Supernova Remnant N103B Using HST}

\correspondingauthor{William P. Blair}
\email{wblair@jhu.edu}

\author[0000-0003-2379-6518]{William P. Blair}
\affiliation{The Henry A. Rowland Department of Physics and Astronomy, 
Johns Hopkins University, 3400 N. Charles Street, Baltimore, MD, 21218, USA; 
wblair@jhu.edu}

\author[0000-0002-9886-0839]{Parviz Ghavamian}
\affiliation{Towson University, Towson MD 21252, USA;
pghavamian@towson.edu}

\author[0000-0002-7868-1622]{John C. Raymond}
\affiliation{Harvard-Smithsonian Center for Astrophysics, 60 Garden Street, Cambridge, MA, 02138, USA;
raymond@cfa.harvard.edu}

\author[0000-0003-2063-381X]{Brian J. Williams}
\affiliation{NASA Goddard Spaceflight Center, Greenbelt, MD 20771, USA;
brian.j.williams@nasa.gov}

\author[0000-0001-8858-1943]{Ravi Sankrit}
\affiliation{Space Telescope Science Institute,
3700 San Martin Drive, Baltimore, MD, 21218, USA; 
rsankrit@stsci.edu}

\author[0000-0002-4134-864X]{Knox S. Long}
\affil{Space Telescope Science Institute,
3700 San Martin Drive,
Baltimore MD 21218, USA;}
\affil{Eureka Scientific, Inc.
2452 Delmer Street, Suite 100,
Oakland, CA 94602-3017, long@stsci.edu}

\author[0000-0001-6311-277X]{P. Frank Winkler}
\affiliation{Department of Physics, Middlebury College, Middlebury, VT, 05753, USA; 
winkler@middlebury.edu}

\author[0000-0003-3382-5941]{Norbert Pirzkal}
\affiliation{Space Telescope Science Institute,
3700 San Martin Drive, Baltimore, MD, 21218, USA; 
npirzkal@stsci.edu}

\author[0000-0002-5044-2988]{Ivo R. Seitenzahl}
\affiliation{School of Science, University of New South Wales,
Northcott Drive, Canberra, ACT, 2600, Australia; 
i.seitenzahl@adfa.edu.au}




\begin{abstract}

We present \hst/WFC3 multiband imagery of N103B, the remnant of a Type Ia 
supernova in the Large Magellanic Cloud, as well as \hst/COS ultraviolet 
spectroscopy of the brightest radiatively shocked region.  The images show a 
wide range of morphology and relative emission-line intensities, from 
smooth Balmer-line dominated collisionless shocks due 
to the primary blast wave, to clumpy radiative shock filaments due to secondary 
shocks in density enhancements.  
The COS data show strong FUV line emission despite a 
moderately high extinction along this line of sight. We use the COS data with previous 
optical spectra to constrain the shock conditions and refine the 
abundance analysis, finding abundances typical of the local interstellar 
medium within the uncertainties. Under an assumption that the material being shocked was shed from the pre-supernova system, this finding places constraints on any significant enrichment in that material, and thus on the non-degenerate star in what was presumably a single-degenerate Type Ia supernova.

\end{abstract}




\section{Introduction} \label{sec:intro}

Type Ia supernovae (SNe) are now widely understood to arise from the thermonuclear destruction of a white dwarf star, but the progenitor systems and exact trigger mechanisms for the explosion are still unclear \citep{heringer17,livio18}.  Type Ia supernovae are believed to be caused either by coalescence of two white dwarf stars 
 \cite[the so-called double-degenerate scenario:][]{iben84,webbink84,livio03},
or by an alternative mechanism, runaway nuclear burning of one type or another on a white dwarf star caused by accretion from a non-degenerate companion star 
\cite[single-degenerate scenario:][]{whelan73,nomoto82a,nomoto82b,nomoto18}.   
Within these two general pathways lie a number of additional theoretical possibilities, such as common envelope scenarios and core-degenerate models \citep{kashi11,Ilkov12,wang12,meng17,wang17}.  These various scenarios have ramifications for the progenitor system, as reviewed recently by \cite{pilar19}.
Many authors have suggested that the double-degenerate pathway dominates \cite[cf.][]{totani08,liu18} but there remain intriguing counterexamples that seem to point toward some version of a single-degenerate scenario for at least some Type Ia SNe \cite[][]{blair07,reynolds07,williams11,williams12}, including N103B---the subject of this paper \citep{williams14,sano18}.

Conclusive evidence in favor of the single-degenerate mechanism has been difficult to find, however.  One of the most promising methods used so far is to search for evidence of interaction between the supernova remnant and a dense circumstellar wind of the type expected to be generated prior to the explosion by the donor star \citep{hachisu96,badenes07,chiotellis12}.\footnote{However, see \cite{levanon15} and \cite{levanon19} for a discussion of possible circumstellar material from certain double-degenerate models.}  One promising candidate is Kepler's supernova remnant, wherein the Balmer-dominated shocks are driving secondary shocks into dense clumps, producing radiative shocks \citep{blair91,sankrit16}.  There is evidence from \spitz\ observations for a non-standard silicate-rich dust component in the shocked material \citep{williams12}, presumably indicating an origin in the mass lost from the stellar wind of the non-degenerate donor star \citep{ossenkopf92,henning10}.  Since Kepler's SNR is well off the plane of the Galaxy, a clumpy, dense surrounding interstellar medium (ISM) is not expected, thus strengthening the conclusion that the shocked material is circumstellar and not interstellar in origin. 

Recently, a second  SNR has been identified as a candidate for a single-degenerate Type Ia:  N103B (SNR 0509$-$68.7) in the Large Magellanic Cloud \citep{williams14}. X-ray analyses of N103B \citep{hughes95,lewis03,lopez11,yamaguchi14} indicate an Fe/O abundance and morphology consistent with a Type Ia explosion.  This conclusion is confirmed by an actual spectrum of the SN derived from a light echo (A. Rest, private communication). In optical images, the SNR exhibits a prominent collection of compact, bright knots emitting in \ha, \sii, and \oiii\ \citep{williams14,Li17} embedded within a smooth partial shell filaments emitting only H Balmer line emission.   \cite{Li17} also identify a potential companion star left behind after the explosion which, if true, would strengthen the case for a single-degenerate explosion.\footnote{To our knowledge, no confirmation of this candidate has been made, although \cite{Li19} have recently identified possible companions in two other LMC SNIa remnants using a kinematic criterion.}  

\spitz\ observations by \cite{williams14} showed very bright 24 $\mu$m emission that arises from warm dust heated by the forward shock, seen in \ha\ and with morphology reminiscent of a bow shock open to the east. Such a structure is consistent with the shock sweeping into a wind from the progenitor system within an overall density gradient, which is similar to the structure seen in Kepler's SNR.  The derived densities in both Kepler and N103B are also nearly identical, $\sim$ 40 - 45 $\rm cm^{-3}$, which would be high for normal ISM ($\le$1 cm$^{-3}$).  The \spitz\ IRS  spectrum of N103B is essentially identical to that of Kepler, with a peculiar `18 $\mu$m' silicate bump offset to $\sim$ 17.2 $\mu$m and a warm dust continuum ($\sim$ 115-130~K; \cite{williams14}).  All of these similarities to Kepler's SNR imply a similar origin for the material being encountered by the shock in N103B, i.e., circumstellar material from the progenitor system.

The size of this SNR ($\sim$30\arcsec\ across, or D = 7.2 pc at the LMC distance of 50 kpc) indicates a SNR larger than Kepler (D = 4.9 pc assuming a distance of 5 kpc) but similar in size to Tycho's SNR (SN 1572).   The light echoes from N103B mentioned earlier place its age at approximately 860 years \citep{rest05}, which would make N103B nearly twice as old as Tycho and Kepler. However, a younger age is possible according to \cite{rest05}, depending on the characteristics of the dust screen responsible for the light echo. \cite{ghavamian17} estimate a Sedov age of 685 $\pm$ 20 years, while an X-ray expansion measurement by \cite{williams18} also implies an age less than 850 years.  Since the distance (and diameter) are known, one can calculate a mean expansion velocity for an assumed age.  Using the age range 685 -- 860 years, the mean expansion velocity must have been in the range {\bf 5009 -- 3986} $\kms$, respectively.  The current average observed shock velocity reported by \citet{ghavamian17} is 2070 $\pm$ 60 $\kms$, well below this range, indicating significant deceleration of the primary shock has occurred.

Recently, \citet{Li17} have presented both narrowband \ha\ and broadband imagery of N103B taken with the WFC3 instrument on \hst\ as well as ground-based echelle longslit spectroscopy.  
The \hst\ images indicated that the filaments forming the outer shell of N103B  exhibit the smooth, delicate morphology characteristic of Balmer-dominated shocks.  Their spectra confirm broad Balmer emission from these filaments, but their observations were not sufficient to reveal the full extent of the broad component  velocities.
  
\citet{ghavamian17}  presented integral field observations of N103B acquired with the Wide Field Imaging Spectrograph \cite[WiFeS;][]{dopita07,dopita10} on the 2.3\,m telescope of the Australian National University.  These observations covered the spectral range 3500 -- 7000 \AA\, and allowed measurement of the full extent of the broad H$\alpha$ line both in the interior and along the southern and northern limbs of N103B.  Coronal [Fe~XIV] $\lambda5303$ emission from the nonradiative shocks was detected in MUSE observations by \cite{ivo19}. \cite{dopita19} presented an in-depth analysis of spectra from a number of LMC SNRs, including the brightest radiative knot near the center of N103B\@.  They performed a detailed assessment of elemental abundances, finding that the abundances of heavy elements in these SNRs are consistent with standard LMC abundances as derived from other sources.   \cite{Li17} also showed high-resolution spectra of this same bright knot in N103B, which revealed highly redshifted ($\sim370 \kms$), somewhat broadened ($\sim190 \kms$) H$\alpha$ emission.

\cite{sano18} have investigated the molecular gas in the direction of N103B using CO observations with ATCA and the Atacama Submillimeter Telescope Experiment (ASTE), covering both moderate and high resolutions.  Interestingly, molecular clouds are seen directly adjacent to N103B, but in the south and southeast, not in the west where the optical nonradiative shocks are brightest.  Hence, while there may be some interaction with molecular gas, it does not appear to be in full force yet; otherwise,  the optical and X-ray shocks would be brighter in the south and southeast.

Here we present additional \hst\ Wide Field Camera 3 (WFC3) narrow band imagery of N103B in the light of selected optical and near-IR emission lines.   
We also present a far-ultraviolet Cosmic Origins Spectrograph (COS) spectrum of the central radiatively shocked knot in N103B (the same cloud targeted by \citet{dopita19} and \cite{Li17}).  
The COS spectrum provides improved abundance information on elements like carbon and silicon that are poorly constrained by optical data alone and can be affected by shock processing of dust grains. 
We describe these observations in the next section, present the modeling and abundance determinations in section 3, and then discuss the results in the context of comparison with Kepler's SNR in section 4.  We summarize our findings in section 5.

\section{Observations  \label{sec:observations}}

The HST data discussed and used below are available from the Mikulski Archive for Space Telescopes (MAST) and can be accessed at the following URL: \url{http://dx.doi.org/10.17909/t9-bddp-1b43}.

\subsection{HST Imaging}

The WFC3 data were obtained on 2017 Jan 3 under program ID GO-14359, which was part of a joint Chandra-\hst\  program (B. J. Williams, PI); both the UVIS and IR cameras were used, as shown in Table\, \ref{hst_obsns}.  The UVIS data included the filters F657N (\ha), F673N  ([S II]), F502N ([O III]) and F547M (continuum), and the IR camera data included F164N ([Fe~II]) and F160W (continuum), thus sampling the bright emissions expected from radiative shocks, with \ha\ also sampling the fainter nonradiative shocks identified by \cite{Li17} and \citet{ghavamian17}, which are dominated by hydrogen Balmer emission.  The continuum bands were nominally for identifying stars and, as needed, subtracting them to provide a cleaner look at the SNR emission.  
The observations obtained in each filter were dithered to permit removal of artifacts from the data and (for UVIS) to cover the chip gap, and the appropriate FLASH parameter was set for UVIS to reduce the effects of charge transfer inefficiency.  The combined images in each filter that were produced by the data processing pipeline were sufficient for the needs of this program.

The F547M filter was selected to obtain a continuum band reasonably close to the bandpass of the other optical filters to identify stars.  However, N103B exhibits a number of faint (and yet unusually strong for typical SNR emission) \feii\ emission lines that lie within the bandpass of F547M \cite[cf.][Table 8]{dopita19},\footnote{While the [Fe~XIV] $\lambda5303$ line seen by \cite{ivo19} is also within the bandpass, the lines strengths listed in \cite{dopita19} indicate \feii\ dominates by at least a factor of 6.} causing the bright radiative filaments to be visible at a low level within the F547M image (see Fig.\, \ref{fig_overview}).  For display purposes, we scale and subtract the F547M data from the emission line frames, but since the stellar density is not severe, flux information has been derived directly from the original emission-line data frames.

\begin{figure}
\plotone{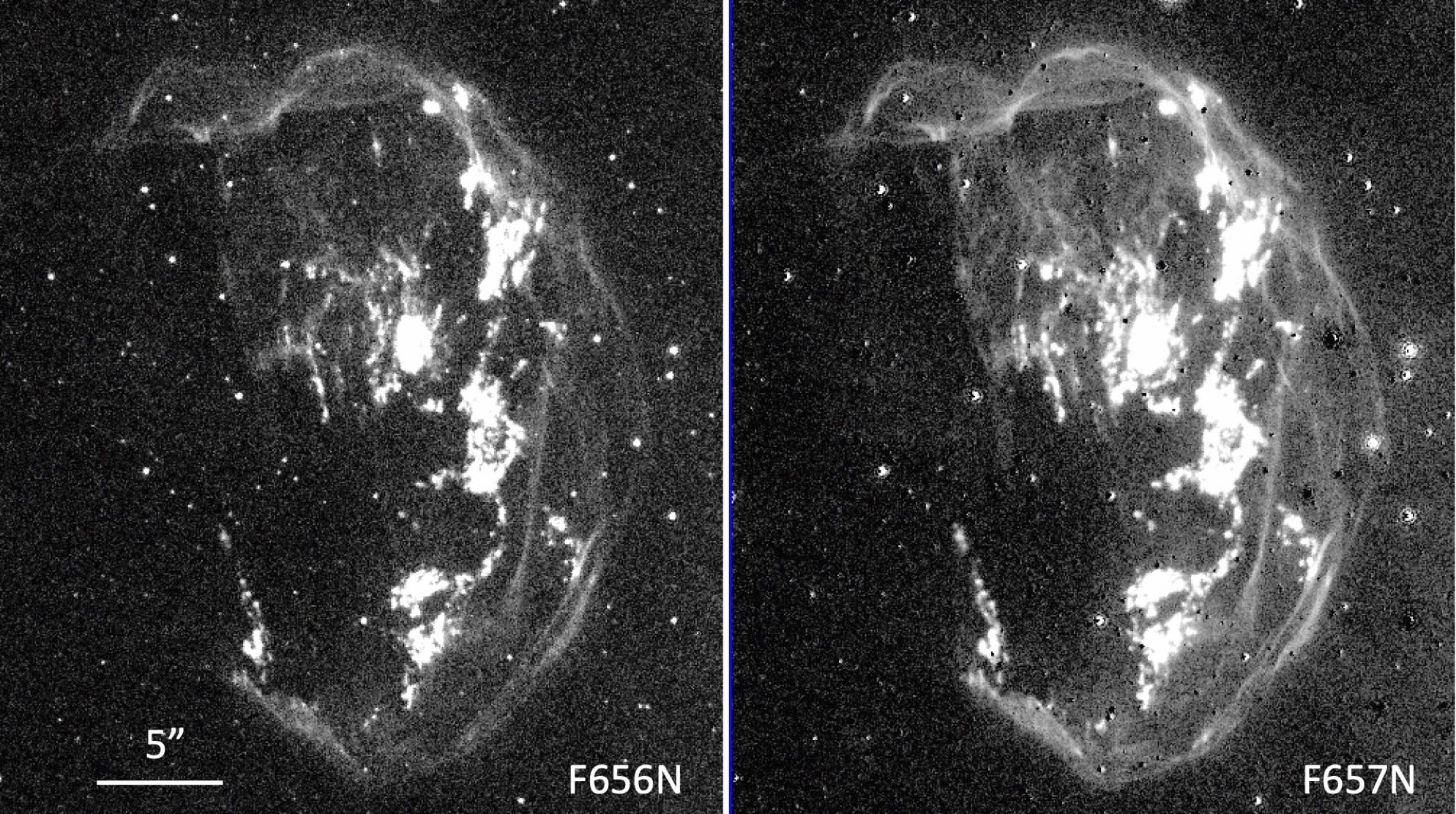}
\caption{A comparison figure showing aligned WFC3 images of N103B in the narrow F656N filter (program 13282) and our wider F657N filter image.  Log scaling of the display is used to bring out the details in the faintest emission.  The broader F657N filter captures significantly more detail in the faint, smooth filaments, which are attributed to nonradiative emission from the primary shock front.  Both images have had a continuum image subtracted to remove stars to first order, but a number of stellar residuals remain.  Close inspection of stellar-appearing features can reveal whether the feature is a real emission knot or a stellar residual from imperfect subtraction.  In this and other images, north is up and east to the left.
\label{fig_hacomp}
}
\end{figure}

\begin{figure}
\plotone{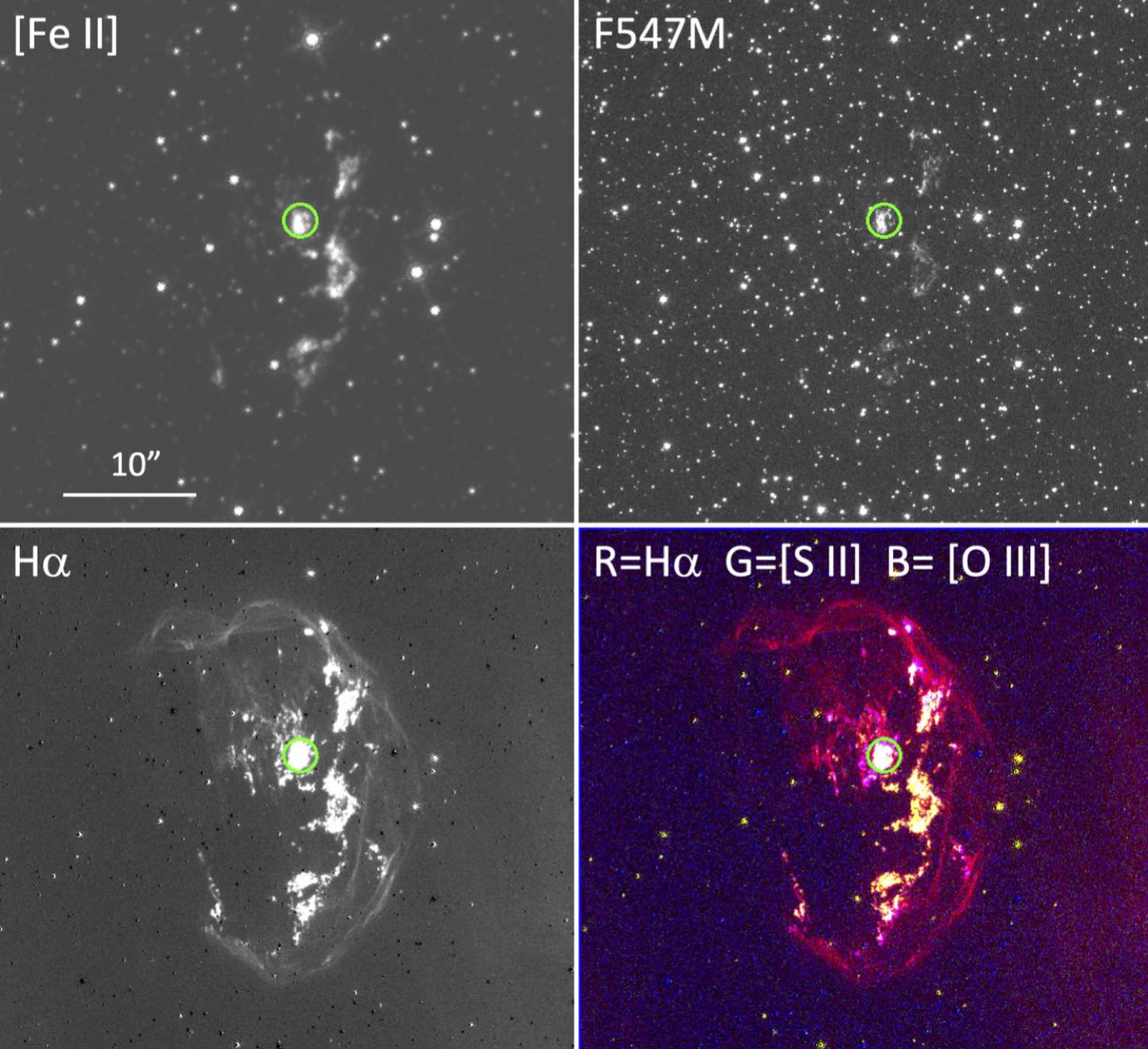}
\caption{A four-panel figure showing aligned images of a 40\arcsec\ region of \hst WFC3 imagery centered on N103B.  The panels show (upper left) \feii\ 1.644 $\mu$m, (upper right) F547M, (lower left) \ha, and (lower right) a color frame with \ha\ in red, \sii\ in green, and \oiii\ in blue. Emission showing in yellow and white are radiative filaments while the red outer shell shows the nonradiative main blast wave.  The green circle shows the position and size of the COS aperture.   
\label{fig_overview}
}
\end{figure}

In this program the wider F657N filter (FWHM = 121 \AA) was used instead of the F656N filter (FWHM = 18 \AA) used by \cite[][program GO-13282, Y.-H. Chu, PI]{Li17} in order to make sure that the full range of \ha\ emission was detected, especially given the $\sim 250 ~ \kms$ redshift of the LMC itself.  Our ``\ha'' image thus contains both emission from \NiiL\ and \ha\ for the radiative filaments, which can be identified by their clumpy, structured morphology. Even so, \nii\ is fairly weak compared with \ha\ in the LMC due to relatively low N abundance, so \ha\ is expected to dominate the F657N image even for the radiative filaments.  We have reprocessed and astrometrically aligned the F656N data set to our new data (registration within 0.2 pixels), and a comparison is shown in Fig.~\ref{fig_hacomp}.  This figure is scaled so as to show the faint emission to best effect, and the improved signal-to-noise for the faint Balmer filaments is apparent.  Details of the structure of faint Balmer filaments can be clearly seen in the right panel of Fig.~\ref{fig_hacomp}.  We note that both images in Fig.~\ref{fig_hacomp} have been continuum subtracted using the F547M image.  Close inspection, especially for the F657N frame, makes it clear that there are a number of very small angular size knots of emission that are real, while stellar residuals can be identified due to a very small offset in alignment between F547M and the emission-line filters.

The  WFC3 IR camera was used to observe \FeiiL\ using the F164N filter, and F160W H-band was used for continuum.  The emission line of interest is actually within the passband of the continuum filter, so some SNR emission is visible in the raw F160W data.  However, the star density is again not severe and \feii\ fluxes can be extracted directly from the F164N image.

\begin{figure}
\plotone{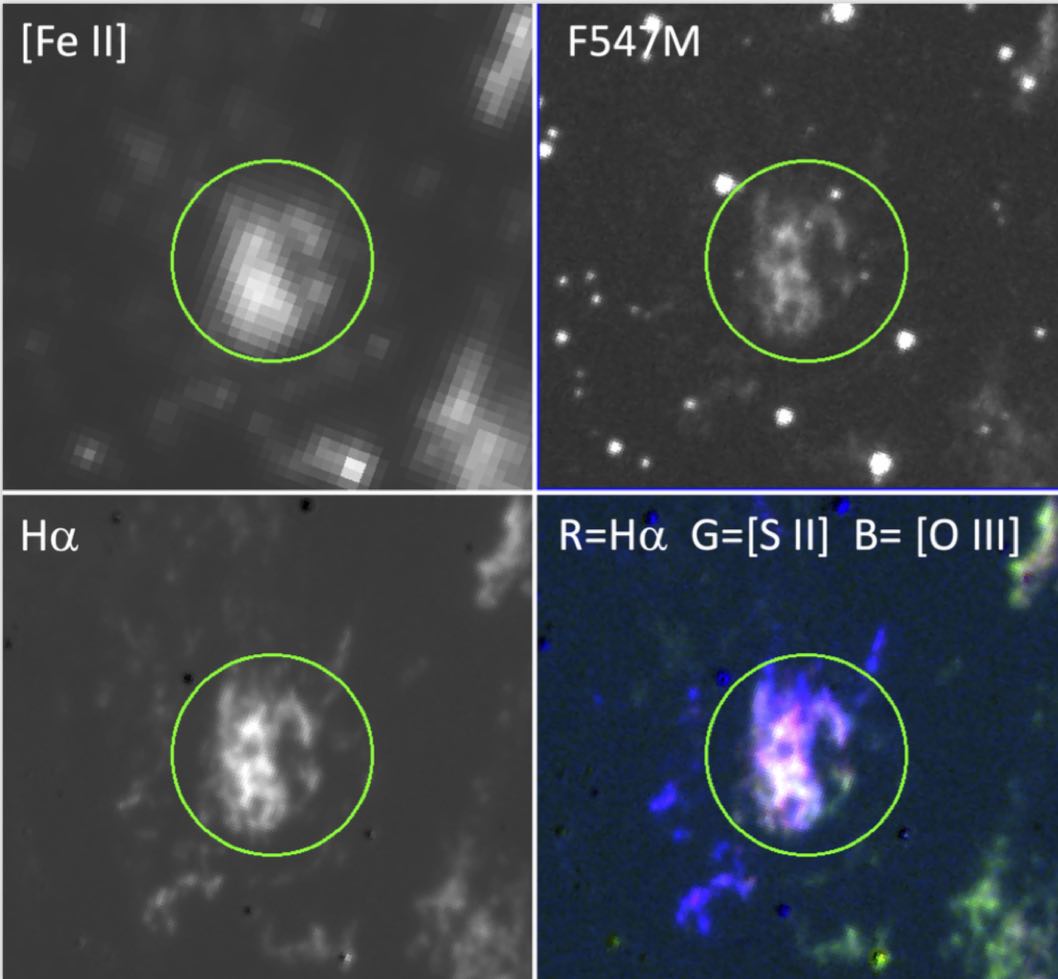}
\caption{A four-panel figure  of a 6\arcsec\ region centered on the COS 2.5\arcsec\  diameter aperture position. The panels are the same as in Fig.\, \ref{fig_overview} but  the scaling has been adjusted to show the detail in the bright filaments.  The faint emission filaments seen in the F547M continuum frame are likely due to faint \feii\ lines that lie within the filter bandpass. They represent a higher spatial resolution version of the brighter \feii\ 1.644 $\mu$m emission seen directly in the upper left panel. Note the complexity of the emission knot sampled by the COS aperture, with different colors indicating varying ratios of the relative line intensities. 
\label{fig_cosim}
}
\end{figure}

We show the resulting images in Fig.~\ref{fig_overview} and Fig.~\ref{fig_cosim}.  Fig.~\ref{fig_overview} shows a set of 40\arcsec\ regions centered on N103B and indicates the full extent of optical/NIR emission.  The color panel is made with the three optical emission lines, after subtracting a scaled version of the F547M image.  Again, for display purposes, no correction for the faint filament emission visible in F547M was made.
The faint, smooth, filament arcs visible only in \ha\ (red) are the fast nonradiative shocks described earlier showing the position of the primary shock front; their full extent and morphology are best seen in the right panel of Fig.~\ref{fig_hacomp}.  The filaments that show as yellow or white in the color panel of Fig.~\ref{fig_overview} are bright in \ha, \sii, and \oiii, and thus represent emission from radiative knots and filaments from slower secondary shocks being driven into density enhancements encountered by the primary shock.   The larger pixel size of the IR camera (0.13\arcsec) compared to UVIS (0.04\arcsec) is obvious in the resolution of  \feii\ panel, but it is clear that we are  seeing \feii\ emission only from the radiative shock regions and not the nonradiative shocks.

\begin{figure}
\plotone{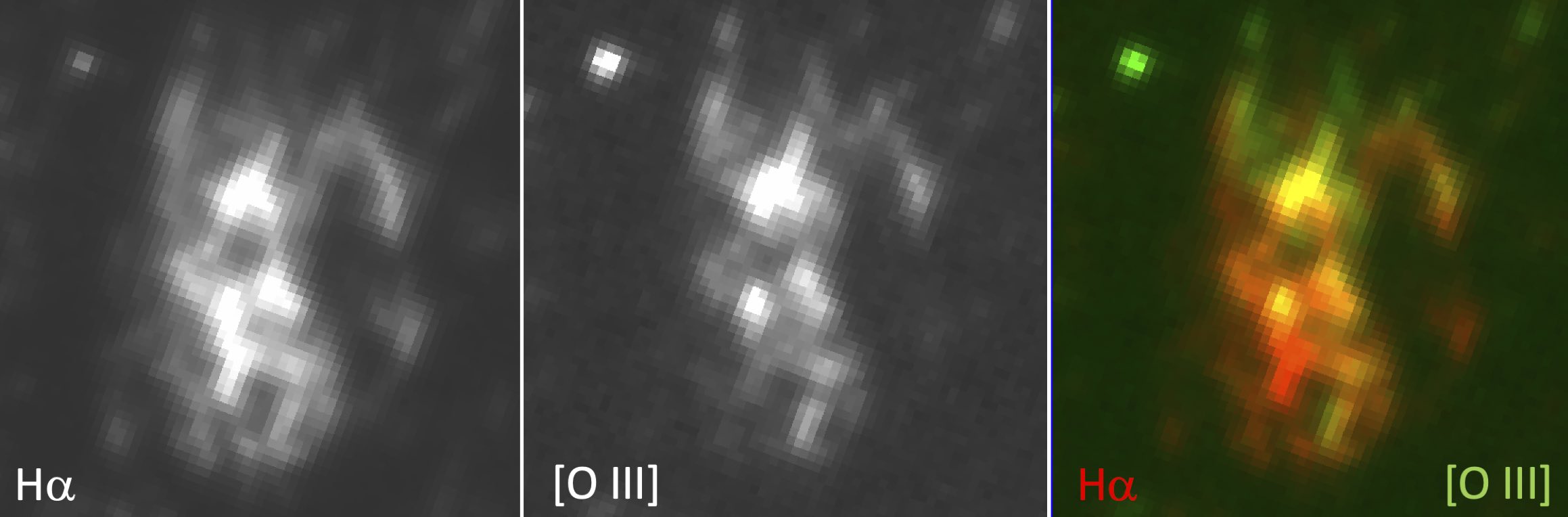}
\caption{A comparison of the sub-arcsecond structure of the bright filament observed with COS, as viewed in \ha\ + \nii\ (marked `\ha') (left), \oiii\ $\lambda5007$ (center), and colorized (right). The region shown is 2.5\arcsec\ square. Using the average line intensities from the optical spectrum, we can convert the ratios observed in these filters to the equivalent \oiii\ $\lambda5007$ to \hb\ ratio, which is less reddening dependent. Regions appearing green (top of Figure) have $\lambda5007$:\hb\ $\sim$2.2 while the bright red region below center has ratio 0.33.  The bright yellow knot above center has $\lambda5007$:\hb\ $\simeq$ 1.4.
\label{fig_zoom}
}
\end{figure}

The green circle in Fig.~\ref{fig_overview} shows the position of the COS aperture, which was used to observe the same bright central knot of emission whose optical spectrum was observed by \citet{dopita19} and \cite{Li17}.  Fig.~\ref{fig_cosim} shows a zoom into this small region, showing the same data as in Fig.~\ref{fig_overview} but with the display scaled to show the detailed structure in the brightest emission.  It is clear from Fig.~\ref{fig_cosim} that the emission sampled by the COS aperture is very complex and likely includes emission from a range of shock velocities and densities. 
Since the emission being passed by the F547M filter is likely dominated by several faint \feii\ lines, this emission represents a higher spatial resolution version of what is being observed at higher signal-to- noise ratio in \FeiiL.  This emission is morphologically almost identical to the \ha\ emission of the knot, demonstrating how closely the \feii\ emission is tracing the radiative shocks in this region.  In Fig. \ref{fig_zoom}, we zoom in one more time, and simplify the display to show only the \oiii\ and \ha\ emission, highlighting the sub-arcsecond variations in ionization structure with the filaments observed with COS.

\begin{figure}
\plotone{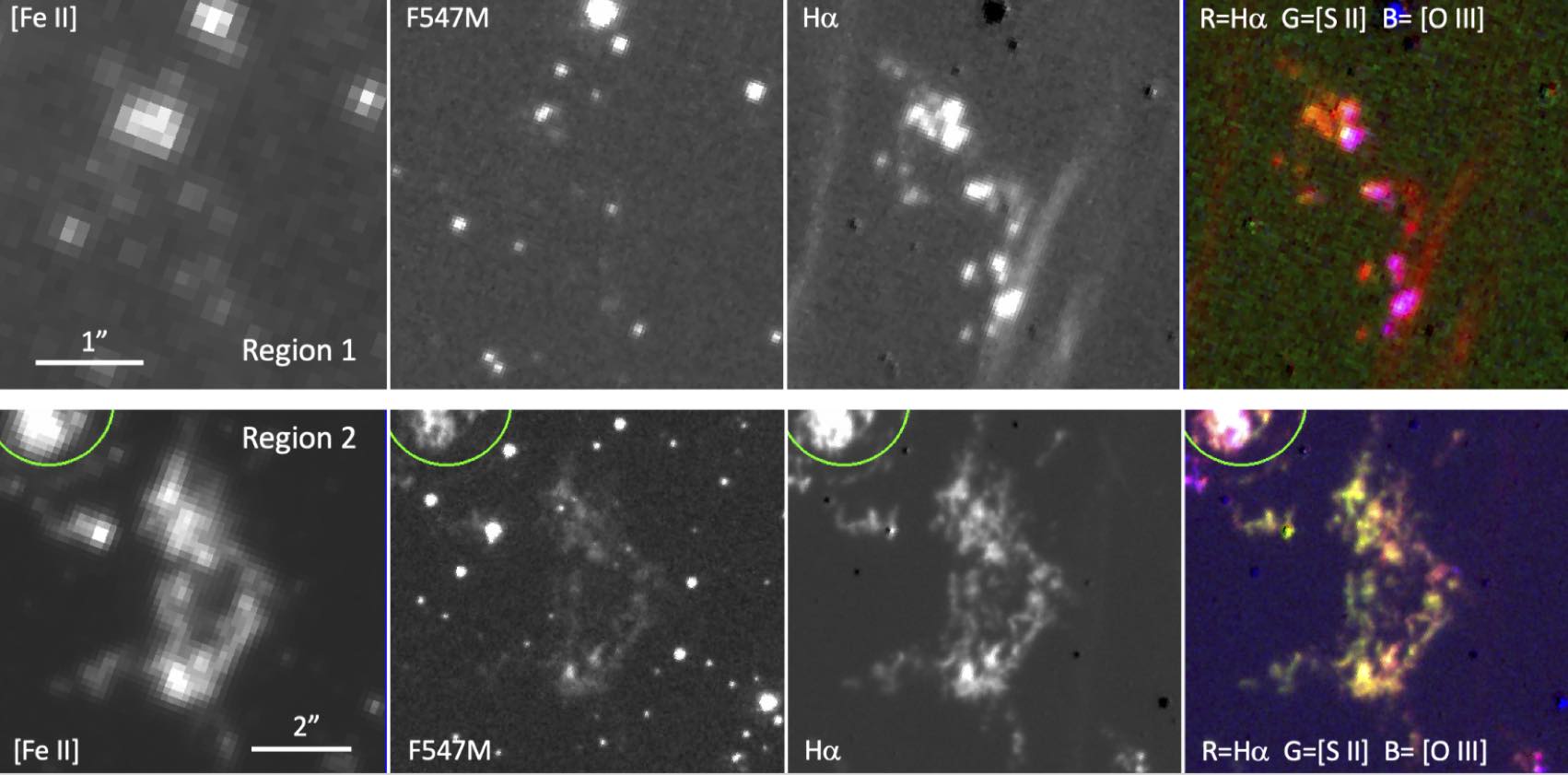}
\caption{This figure shows two regions of clumpy, radiative filaments in N103B that have very different emission properties.  The top four panels show Region 1, a 4\arcsec\ region centered $\sim$9\arcsec\ SW of the bright central knot, near the SW limb of the SNR. It shows knotty filaments primarily seen in \ha; minor admixtures of \oiii\ (magenta) and/or \sii\ (orange) indicate these are knotty nonradiative filaments in the early stages of transition to becoming radiative.  The bottom four panels show Region 2, a 7\arcsec\ region of brighter radiative filaments just a few arcsec to the SW of the COS aperture position. These filaments are well on their way to establishing full cooling and recombination zones, but the relative line strengths vary, causing the color variations seen.  Note how the \feii\ emission has `turned on' for these filaments.  This same complexity is happening within the region observed with COS but on a more compressed spatial scale.
\label{fig_radiative}
}
\end{figure}

In general, the character and morphology of the optical filaments in N103B  are very reminiscent of what is seen in Kepler's SNR  \citep{sankrit08}, a similarity that has been noted previously \citep{williams14}.  Fig.~\ref{fig_radiative} shows two additional examples of small regions of radiative filaments,  enlarged to show their detailed structures.  The top panels show Region 1, a small region of knotty filaments on the SW limb, seen (at least in projection) within some of the smooth, faint nonradiative filaments from the main blast wave.  In the color panel, these filaments are primarily red, meaning they are strong in \ha.  However, there are subtle color differences, with magenta filaments including some \oiii\ emission and a couple of orange knots toward the north, meaning some \sii\ emission is present as well.  These are presumably density enhancements that have been struck by the main blast wave relatively recently and are in the early, incomplete stages of transition to becoming radiative.   The northern (orange) filaments are the only ones that show significant \feii\ emission, consistent with the idea that these filaments are farther along toward becoming fully radiative. Again, the presence of these transition filaments is very similar to what has been seen in Kepler's SNR \citep{blair91}, the only other SNR for which this morphology has been observed.

The bottom panels of Fig.~\ref{fig_radiative} show Region 2, a grouping of well-developed radiative filaments just to the SW of the COS position (that is, seen in projection much closer to the center of the remnant).  \cite{Li17} refer to these filaments as their grouping III.  The color variations seen in the rightmost panel show that the relative fluxes of the various emission lines are changing, but it is not clear how much of this may be due to changing physical conditions (variable density and hence effective shock velocity) and how much may be due to shock incompleteness effects (variable time since a given filament encountered the primary shock).  The same complexity is likely occurring in the bright feature observed with COS, but on a compressed spatial scale, at least as seen in projection.

\subsection{COS FUV Spectroscopy}

The 2.5\arcsec\ point source aperture of COS was placed on the bright central knot in N103B as shown in Fig.~\ref{fig_overview}.  Table \ref{hst_obsns} includes the observational details.  The COS G140L grating was used at all four FP-POS positions to produce a spectrum covering the range 900-2000 \AA, although the data below $\sim$1150 \AA\ and above 1950 \AA\ are too noisy for use.   The data were processed with the standard pipeline processing via CalCOS v2.12. These data were obtained at COS lifetime position 3, for which there are some complications for extended sources that fill the 2.5\arcsec\ aperture; some flux may be lost at locations where the extended source emission overlaps a low-sensitivity portion of the detector.  However, while Fig.~\ref{fig_cosim} shows the bright knot to be resolved, the bright emission does not fill the aperture entirely.  Our assessment of relative line intensities below seems to indicate the impact is small, but this adds some uncertainty to the interpretation of relative line intensities.

\begin{figure}
\plotone{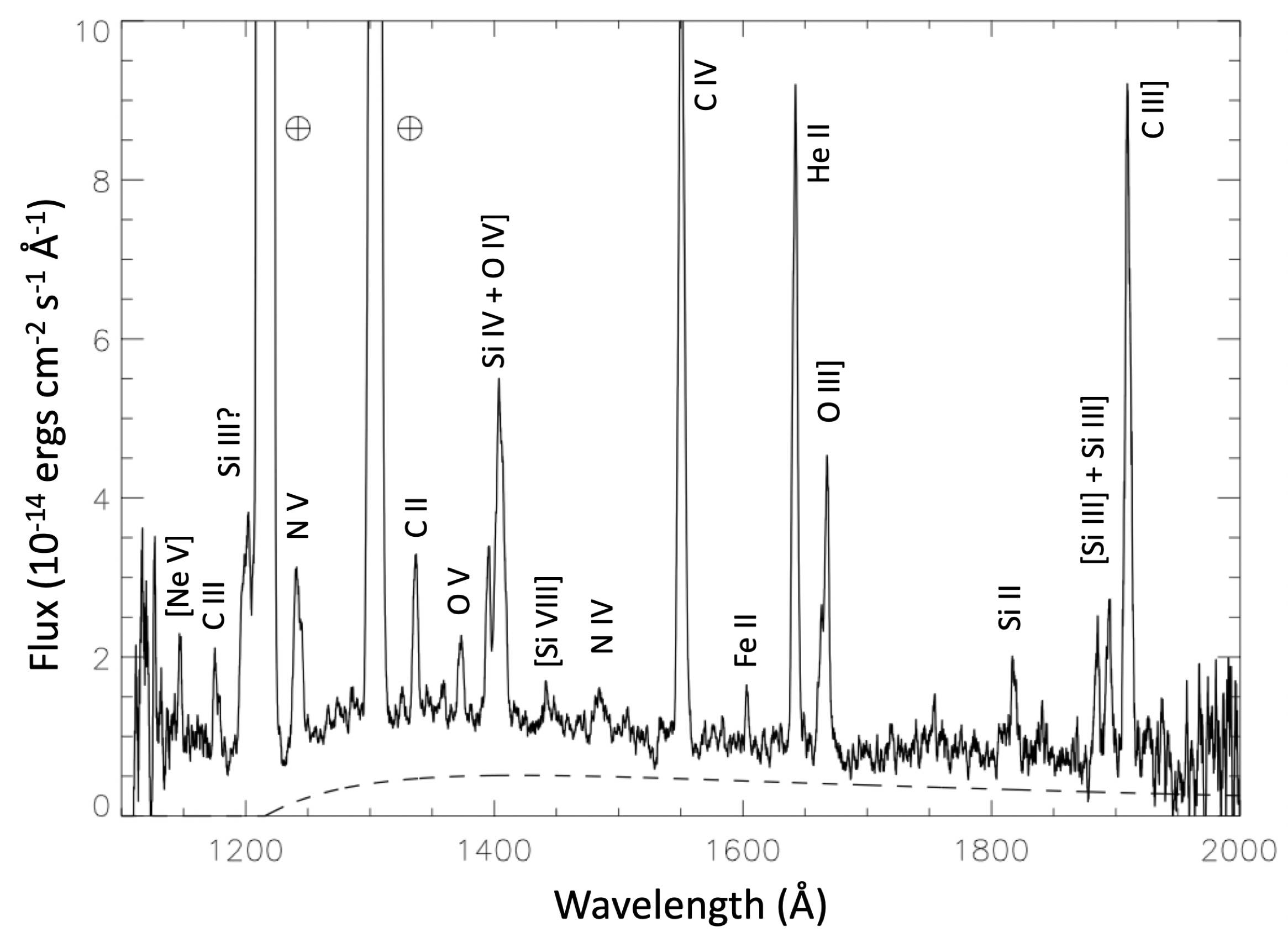}
\caption{The COS spectrum is shown after correction for reddening as described in the text.  The line identifications are shown.  The dashed line indicates the shape of the 2-photon continuum spectrum for reference; this was calculated for a 250 $\kms$ shock.  However, neither the shape nor the strength of the continuum is well-matched by two-photon alone, implying blue starlight is present at a low level.  See text for details.
\label{fig_cosspec}
}
\end{figure}

The FUV line intensities and line widths have been measured using a python-based Gaussian fitting routine.  Values and 1$\sigma$ errors are provided in Table \ref{cos_obsns}. We apply the \citet{fitz86} average LMC extinction curve with a value of E(B - V) = 0.33 (determined from the optical Balmer lines -- see below) to deredden the spectrum.\footnote{\cite{lewis03} report an N(H) absorption column in X-ray of 3 -- 4 $\rm \times 10^{21} ~ cm^{-2}$, consistent with a moderately high extinction on this line of sight.}  The derived intrinsic fluxes are also shown in  Table \ref{cos_obsns}, both in physical units and scaled relative to [O~III] $\lambda$1666 = 100.  These scaled intensities will be used to compare with model calculations below.  In Fig.~\ref{fig_cosspec}, we show a version of the spectrum with the extinction correction applied that also shows the line identifications.

Because of the extended nature of the emission within the aperture, the spectral resolution achieved is lower than that expected for a point source.  Based on measurements of the strong \oi\ airglow line at 1356 \AA, we find the width of an emission line filling the COS aperture to be 7.9 \AA.  While emission certainly fills the COS aperture, the brightest emission is more compact and hence observed SNR lines are somewhat narrower, depending on the internal kinematics of the emitting material. For example, He~II $\lambda$1640 is a strong, single line and shows a FWHM of 4.4 \AA, or about 800 $\kms$.  With shock velocities near 200 $\kms$ derived below, it is clear we are not resolving actual kinematic information from these low-resolution data.  

A low-level but significant continuum is present in the COS spectrum.  Some fraction of this component may be due to hydrogen two-photon continuum, as shown by the dashed line in Fig.~\ref{fig_cosspec} (see discussion below), but the continuum peaks at shorter wavelengths than expected for the two-photon continuum,  and it extends to the short wavelength side of Ly$\alpha$, possibly even showing a broad Ly$\alpha$ absorption line.  It seems likely that a significant fraction of this continuum is simply scattered star light from the general region of the LMC bar, as noted in the far-UV spectrum of N103B from FUSE \cite[see][Appendix sec.~A6 and Fig.~7]{blair06}, although the aperture in that case was significantly larger.  A close inspection of the F547M panel in Fig.~\ref{fig_cosim} shows a few faint spoiler stars within the aperture that could also be responsible for this faint continuum.

\section{Analysis\label{sec:analysis}}

\subsection{Additional Information on the Bright Knot's Emission}

\cite{dopita19} provided a full set of optical extinction-corrected line intensities for the central knot in N103B (see their Table 8).  However, not knowing exactly what region was extracted for their spectrum, we 
have instead re-extracted some of the key observed optical line intensities from the WiFeS data (Proposal ID: 4140118, PI Seitenzahl) for use here.  Using QFitsView \citep{ott12}, we inspected the WiFeS data cube and selected a 3 $\times$ 3 spaxel region that not only approximated the COS aperture size but also encapsulated the emission of the bright knot with a minimum of contamination from nearby emission.  WiFeS spaxels are essentially 1\arcsec\ square, so the listed optical fluxes are effectively surface brightnesses per square arcsec averaged over the 9 spaxels that were extracted.  We use the observed ratio of the Balmer lines with the \cite{cardelli89} extinction curve to determine E(B - V) = 0.33 (assuming R=3.1), and then apply this correction to the observed optical line intensities to obtain intrinsic relative intensities, as shown in Table \ref{cos_obsns}.\footnote{The LMC and galactic extinction curves are very similar over the optical wavelength range, but different in the UV due to a larger number of small grains in the LMC; hence, the use of the \cite{fitz86} curve for the COS spectrum.}  The relative scaling between the optical and COS FUV lines is complicated by a number of factors, such as the very different spatial resolution and somewhat different aperture sizes as well as the large separation in wavelength.  Therefore, we choose instead to work with relative line intensities within each wavelength range as much as possible.  In the bottom section of Table \ref{cos_obsns}, we show the optical lines for the COS knot scaled relative to I(\hb) = 100 for comparison to the models.

We also extracted the fluxes within the projected COS aperture from each of the WFC3 images.  Besides providing a simple sanity check on fluxes, this allows us to scale the \FeiiL\ line intensity into the comparison in a reasonable way.  The IRAF\footnote{IRAF is distributed by the National Optical Astronomy Observatory, which is operated by the Association of Universities for Research in Astronomy, Inc., under cooperative agreement with the National Science Foundation.} task `imexamine' was used to derive total counts in the projected COS aperture from each WFC3 filter image.  These counts were then converted to fluxes using conversion factors derived from the file headers. We find F(\feii) = $9.22\times 10^{-14} ~ \FLUX$; F(\oiii) = $9.85\times 10^{-14} ~ \FLUX$; F(\sii) = $9.58\times 10^{-14} ~ \FLUX$; and F(\ha) = $5.68\times 10^{-13} ~ \FLUX$.  The \feii\ image flux was scaled to the \oiii\ image flux (since it is a single line), and then scaled to the optical relative line intensities to obtain the relative scaling shown in Table \ref{cos_obsns}.

According to \citet{Li17} the emission from the COS position is redshifted by nearly
$400 \kms$ (which is $\sim 150 \kms$ with respect to the LMC local ISM).  This means that the resonance lines C~IV, Si~IV,
N~V, and C~II are nearly unaffected by resonance scattering in the ISM of either the Milky Way or the LMC, which would otherwise  cut into the short-wavelength sides of their profiles.  This is consistent with the results of the Gaussian fits to the line profiles, which imply little asymmetry in the lines.
From Fig.~8 of \citet{Li17}, we estimate that no more than 10 - 15\% of the resonance line flux can be missing.  Hence, given other uncertainties, it is reasonable to use the extinction-corrected FUV line intensities with the WiFeS optical observations of the filament to compare to shock-model calculations.

\cite{sano18} detect two small clumps of CO emission (see their Fig.~7) in addition to the larger molecular cloud emission to the south and east of N103B. Their `clump A' is centered $\sim$3\arcsec\ northeast of the COS aperture and the contours partially overlap the bright knot observed with COS.  They measure a physical size of this clump as $\sim$1.2 pc and estimate a mass of over 100 $\MSOL$.  However, the velocity of this emission is consistent with the LMC rest frame ($\sim246 ~ \kms$) and not with the velocity of the bright knot (centered at 385 $\kms$).  We do not consider this a physical association but rather a projection effect.  Likewise, while their `clump B' overlaps the southwest limb of the SNR, we see no evidence of deformation of the shock front at that location that would indicate a physical interaction with the CO-emitting clump.

\subsection{Comparison to Model Calculations}

In order to interpret the combined COS and optical/IR data, we have calculated shock models using the code first described by \citet{raymond79}, which has been continually updated with improved atomic physics and functionality over the years \cite[cf.][and references therein]{cox85,hartigan04,koo16}. 
The line width observed by \cite{Li17} in the broad \ha\  component of the brightest nebular knots is 190 $\kms$, which is consistent with shock speeds of 200 $\kms$ or more. A number of preliminary models were calculated to constrain the relevant range of preshock density and magnetic field parameters that reproduced the approximate observed postshock density-sensitive line ratios, which all indicate very high densities.  The optical \SiiL\ ratio is in the high-density limit, and the \sii:\ha\ ratio is 0.22 which is low for typical radiative SNR shocks;  this indicates that significant collisional de-excitation of the \sii\ lines is responsible for the low observed ratio.  
But other diagnostics are available \citep{dere97}, such as the Si~III ratio I($\lambda$1885)/I($\lambda$1894) and C~III ratio I($\lambda$1176)/I($\lambda$1909), which indicate densities of $\rm 3 - 10 \times 10^{4}\: cm^{-3}$.\footnote{Interestingly, the Si~III lines are adjacent in wavelength while the C~III lines are widely separated.  Hence, another way to look at this is that the consistency for the density estimates from these two line ratios argues that the extinction correction is approximately correct.} The optical \sii\ $\lambda$4075/$\lambda$6725 ratio also indicates that a high density is needed to match the observations. Postshock densities can be varied in the models either by increasing the assumed preshock  density, or by adjusting the assumed magnetic field, which adjusts the compression in the postshock flow. (Lower magnetic fields allow more compression and thus higher densities at a given location in the post-shock flow.)

In Table \ref{cos_obsns}, results of three models are shown, where the FUV lines are scaled (as with the observations) to O~III] $\lambda$1666 = 100 and the optical lines are scaled to \hb\ = 100.
Models A, B, and C were calculated assuming shock velocities of 200, 220, and 250 $\kms$, preshock densities 3100, 1370, and 1000 $\rm cm^{-3}$, magnetic field strengths 180, 300, and 180 $\mu$G, respectively, and fully preionized preshock gas.    We used the abundances that \citet{dopita19} derived for N103B as a starting point for these models, viz. He, C, N, O, Ne, Mg, Si, S, Ar, Ca, Fe, Ni = 10.92, 8.09, 7.17, 8.37, 7.57, 7.19, 7.11, 6.88, 5.79, 6.02, 7.33, and 5.91, respectively.  A photoionization precursor may contribute to some of the lines, but the contribution is modest \citep{vancura92}, and following \citet{dopita19} we do not include it.

Model B is intended to be similar to the model calculated with the MAPPINGS code by \cite{dopita19} for N103B, with a shock speed of $220 \kms$ and a ram pressure of $\rm 1.33 \times 10^{-6} ~ dyne~cm^{-2}$; they assumed equipartition of thermal and magnetic pressure ahead of shock. 
The abundances seem to work well overall and their derived parameters produce a reasonable fit to the data, especially since it is for a single shock velocity.  Most lines are within 20\% of the observed relative fluxes.  
The exceptions are none other than the two density-sensitive ratios mentioned above, both of which indicate higher densities than given in
Model B by at least a factor of 2.  Our models A and C use a lower magnetic field strength in part to address this discrepancy.  At T $\simeq$ 30,000 K in the cooling zone of the models, Model A has $n_{\rm e}  =  1.2 \times  10^{5} \: {\rm cm}^{-3}$, Model B has $n_{\rm e}  =  2.5 \times 10^{4}\: {\rm cm}^{-3}$, and Model C has $n_{\rm e}  =  2.9 \times  10^{4} \: {\rm cm}^{-3}$,
 giving Si~III ratios of 0.49, 1.48,  and 1.0, respectively, bracketing the  observed ratio of 0.70.

Also, recall that the observed complexity of the bright knot (cf.\ Fig.~\ref{fig_cosim}) makes it likely that a range of shock conditions is actually present within the aperture.  However, the brightest radiative shocks near 220 $\kms$ apparently dominate the observed emission, making even single-velocity models in this range acceptable to constrain abundances, as was recently found for the Cygnus Loop \citep{raymond20}.  
Some of the weaker lines from the highest ionization states hint at more complexity, but of course their observed strengths are also more uncertain.
The weak [Si~VIII] $\lambda$1442 line nominally requires a faster shock, but that would produce too much flux in the He~II $\lambda$1640 and $\lambda$4686 lines.  Given the weakness of the [Fe~VII] optical line, it is likely that grain destruction is incomplete at the temperatures where Si~VIII and Fe~VII are formed, so the range of shock speeds must include speeds above 250 $\kms$. 

\cite{dopita19} had no solid diagnostics for the C and Si abundances, but adopted reasonable values from other published work. The observed Si~IV $\lambda$1394 line is stronger than the models, while the Si~III  $\lambda\lambda$1885,1892 lines agree reasonably well. (The identification of Si~III $\lambda$1206 is considered uncertain due to the strong Ly$\alpha$ airglow feature.)  C~IV $\lambda$1550 is over-predicted by the models, while C~III $\lambda$1909 line is approximately correct in the models that reproduce the $\lambda$1176/$\lambda$1909 ratio.\footnote{Recall that the uncertainty in the weak $\lambda$1176 line and the reddening correction both impact this comparison, however.}  C~II $\lambda$1335 is predicted to be too strong, but it is such a strong resonance line that, of all the permitted transitions, it might be somewhat attenuated by ISM scattering.  Overall, this suggests that the assumed abundances for those elements are approximately correct if grains are completely destroyed by the time the gas cools to $10^5$ K.  We note that the \feii\ 1.644$\mu$m line is substantially overpredicted, suggesting the possibility of a lower Fe abundance or incomplete Fe grain destruction relative to C and Si, though the collisional excitation rates for [Fe II] lines are somewhat uncertain.  Shocks at these speeds are expected to return about half of the iron to the gas phase \citep{dopita16}.  Also, the models are allowed to cool to 1000 K, and \feii\ forms in the final stages of that cooling.  Since it takes almost 700 years for the gas to reach 1000 K, and since the bright radiative shocks have effective ages much less than the age of the SNR, incomplete cooling is very likely the cause of this discrepancy.

The assumed N:O abundance ratio gives an excellent match to the N~V and N~IV] lines relative to the O lines, which confirms the finding of \cite{dopita19} of an ordinary ISM abundance of N.
We  have only an upper limit on H from the 2-photon continuum, but if H had been burned to He the $\lambda$1640 line would be much stronger than observed.
Therefore, it seems that there is no evidence for nuclear burning of the material being encountered by the shock.  The abundances in the bright knot thus look to be entirely consistent with expectations from the interstellar gas in the LMC.

Finally, although we concentrate here  on the FUV spectrum, we note one glaring discrepancy with the predicted optical lines: the \OiiiL\ lines  are  predicted to be too bright by a factor of 2.  This appears to be a generic problem in modeling. \cite{dopita16} modeled the spectrum of the LMC remnant N49 with a single shock of 250 $\kms$ and found that the same factor-of-2 discrepancy in \oiii\ appeared in an otherwise reasonable model match.  This discrepancy may be an indication of thermally unstable cooling \cite[e.g.][]{innes92,sutherland93}, since unstable cooling tends to shift energy from higher to lower temperature regions.  It might also indicate that slower shocks, as seen in the regions of high H$\alpha$ to [O III] ratios in Figure~\ref{fig_cosim}, make a substantial contribution.

Given the complexity of the filament being observed with COS, it is quite surprising how well a single velocity model (or at least a modest range of shock velocity models) can reproduce the observed spectrum.  Referring to Fig. \ref{fig_zoom}, it is possible that the brightest optical filaments dominate the lower ionization lines while the fainter (but higher ionization) filaments seen primarily in \oiii\ actually dominate the FUV emission; when viewed without the benefit of the full spatial resolution of \hst, the combined spectrum approximates a fairly fast and complete shock reasonably well.  But clearly this is an oversimplification of a more complex interaction between the shock and the bright knot at the COS position.

Overall, we find that the abundances used by \cite{dopita19} work well, within the combined uncertainties of the observations and the models.  However, \cite{dopita19} do not consider how much variation in abundances could be present while still providing a reasonable match to their model.  Abundance uncertainties are difficult to assess quantitatively due to the way various parameters besides abundances  can interact in the models that predict the line ratios, but the presence of multiple ionization stages of various ions can at least bound the problem to some degree.  In Appendix \ref{sec:appendix}, we provide a more detailed discussion of this process, and derive the following constraints.
If we adopt measurement uncertainties of 25\% in the UV line ratios and add these in quadrature with the uncertainties discussed in Appendix \ref{sec:appendix}, we arrive at the following abundance estimates and ranges relative to \cite{dopita19} (represented by D19):  N:O =  D19 $\times$ $1.1 \pm\ 0.35$; C:O  =  D19 $\times$ $0.8 \pm\ 0.47$; Si:O =  D19 $\times$ $0.5 \pm\ 0.40$; and from the optical, N:H = D19 $\times$ $1.0 \pm\ 0.5$.  Note that this estimate would allow some modest amount of enhancement in N:H over ISM abundances, although this is not required by the observations.

\subsection{Geometry of the Observed Filament}

The observed redshift of the filament relative to the LMC velocity, as observed by \cite{Li17}, must result from some combination of Doppler velocity and line-of-sight projection angle, $\theta$.  The range of velocities they see then could result from a range of these parameters within the sampled material, which seems likely given the complex morphology of the filament.  Two possibilities occur for explaining the overall redshift.
Either the cloud was moving at fairly high velocity before the shock arrived, or the cloud is seen fairly face-on (but moving away from us) rather than edge-on and has been accelerated by the shock.  The former case would suggest ejection from the progenitor system, while the latter would require a range of shock speeds.

We can eliminate the possibility that the cloud itself was moving at $\sim 200\, \kms$ before
the blast wave hit it in the following manner.  As seen in N49 by \cite{vancura92}, there should be a photoionization precursor visible as a narrow component of \ha.  From \cite{Li17}, it appears that there is a narrow 
component at the LMC rest velocity, but there is no narrow component emission offset $200\,\kms$
(their Figure 8) corresponding to the bright knot.  Adopting the latter approach, the peak is about $150\, \kms$ redward of the narrow LMC component.  Taken at face value, it means that we are looking at the cloud a little ways from face-on (albeit from the rear since it is moving away from us), with cos($\theta$) $\sim 0.6$.   As the gas cools, it is moving at the shock speed, so a $300\,\kms$ shock is enough to account for this velocity.  This geometry is consistent with the lack of apparent resonance-line scattering, which might have been expected from shocks being viewed closer to edge-on. From the velocity perspective, this geometry means that the
highest velocity emission in the profile, which is seen at $\sim 540\, \kms$, or $300\, \kms$ 
redward of the LMC rest velocity, requires some shocks  up to $\sim 500\, \kms$ to be present.

\section{Discussion and Comparison with Kepler's SNR}

Despite the larger age and diameter of N103B, the similarities between it and Kepler's SNR in our Galaxy are truly remarkable, as discussed by \cite{williams14}.  Both show a primary shock front demarcated by Balmer filaments and clumpy radiative shock emission from secondary shocks driven into density enhancements.  Both are expanding into surrounding material characterized by high densities and a density gradient;  most other remnants of SN Ia's are expanding into very low density regions \cite[cf.][Table 3]{williams14}. The Spitzer IRS spectra of the two objects are nearly identical, showing warm dust heated by the primary shock and an anomalous broad silicate feature centered near 17.2 $\mu$m, the latter taken as evidence of circumstellar dust in the outflow from the pre-supernova system \citep{henning10}. This conclusion seems particularly strong for Kepler, whose position hundreds of parsecs off the galactic plane makes a circumstellar origin of the material highly favored.  It is difficult not to conclude the same thing for N103B.  This is our favored explanation.

Our findings in this paper from analysis of both optical and FUV spectral data of N103B have only solidified the view that the material being encountered by the shock has chemical abundances consistent with those expected for the local ISM.  Interestingly, this confirms yet another similarity with Kepler.   \cite{dopita19} have now shown quite conclusively that the abundances in the radiative knots of Kepler's SNR are also consistent with those expected for the local ISM abundances at its location only $\sim$3 kpc from the galactic center, a point that was first noted in \cite{blair07}.
We mention this because there continues to be confusion on this point even in recent literature for Kepler \cite[cf.][]{Li17,sano18,millard20}. 
The confusion is caused by the high observed N abundance in the shocked material in Kepler \citep{blair91}, which over the years has been attributed incorrectly to an enhancement in the circumstellar medium (CSM) from the progenitor system instead of it being a reflection of the abundance gradient in our Galaxy \citep{rolleston00,rudolph06}. 

So what are the implications of the normal ISM abundances in the dense, knotty material around both N103B and Kepler?  Either a) the material is truly ISM material, or b) it is circumstellar (i.e., from the progenitor system) but with abundances consistent with normal ISM. \cite{dopita19} adopt the former conclusion, which ignores the significant evidence pointing to a circumstellar origin of the material being encountered by the shock, especially for Kepler. We note that N103B is located in the outskirts of a large shell of emission that surrounds the nearby cluster NGC~1850.  As \cite{sano18} point out, there are molecular clouds in the vicinity, and their Fig.~1 shows the presence of very faint, diffuse \ha\ emission in the wider field surrounding N103B \cite[see also Fig.~1 in][for an even wider view]{Li17}.  Hence, it is perhaps less compelling that the material surrounding N103B must have originated from the progenitor system. Were it not for the many similarities to Kepler's SNR and its probable CSM interaction, this might be the preferred conclusion.  

Adopting conclusion b) above, it implies that the material lost in the pre-SN wind phase for both Kepler and N103B was not heavily enriched or processed by the precursor system.  This conclusion has obvious implications for the precursor system.  For example, single degenerate models where the companion was a red giant (RG) or asymptotic giant branch (AGB) star would not be favored, as the CSM produced from those systems would be expected to have enhanced N over the ISM abundances \citep{chiotellis12}.  
Searches for the surviving companion in Kepler have largely ruled out brighter possibilities such as RG and AGB companions \citep{kerzendorf14,pilar18}, but possible surviving companions from fainter main sequence or even sub-dwarf or white dwarf companions have been proposed but not yet ruled out \citep{marietta00,shen17,meng19}.  A number of models for single degenerate SNIa exist that involve main sequence companions \citep{han06,meng10} or common envelope phases \citep{kashi11,meng17} that can lead to a significant mass loss but avoid the need for enhanced abundances in the resulting CSM, but these scenarios have not been explored in detail.  It is also likely even in the CSM scenario that some significant amount of ISM will have been swept up and mixed in with the CSM by the time these remnants have grown to  several parsecs in radius, thus diluting any abundance anomalies.

One other significant clue might lie in the mass of the material in the purported CSM component.  In their analysis for Kepler, \cite{dopita19} 
use the observed densities and physical size of the bright western radiative filament grouping in Kepler to estimate that it likely contains several $\rm  M_\sun$ of material.   \cite{blair07} estimated nearly a solar mass of material was involved in the currently heated material in the shell of Kepler. 
For N103B, \cite{williams14} use X-ray and IR data to estimate $\sim$3 $\rm  M_\sun$ of material currently emitting.  For the bright COS filament, the models we have presented show that the assumed preshock density and magnetic field strength play off against each other, but reasonable assumptions about the size, depth, and density can be used to arrive at an estimate of order 0.1 $\rm M_\sun$. \citet{Li17} performed a similar experiment using most of the bright filaments and concluding at least 1 $\rm  M_\sun$ of material is present (with even more if previously shocked material has already cooled).  So overall this seems comparable to the situation in Kepler.  These estimates thus require a significant amount of mass loss from the progenitor system if the CSM option is to be believed. Some of the models investigated by \cite{meng10} have appropriately massive companions that could provide this material if indeed sufficient mass can be lost in the pre-SN phase.  Alternatively, if this is really ISM, one needs to account for why such a dense, structured ISM exists around these SNRs.

While we favor the CSM interpretation, we are unfortunately left with a conundrum as to the true origin of the material being encountered by the expanding blast wave in N103B.  The best way to potentially resolve this issue is to determine whether a fainter surviving companion can be identified.  This would then inform the discussion of what the progenitor system might have been and thus what any material lost from that system might have been like. The star indicated by \cite{Li17} is certainly not an ironclad identification of a companion, but follow-up work is needed.  Also, assuming fainter companions are in play, there are numerous other stars near the center of N103B that were not considered by \cite{Li17}.  Any  search for a surviving companion should reconsider the assumed position of the explosion in determining the region to be searched. \cite{Li17} selected the center of an ellipse fit to the partial \ha\ shell as their center position. From consideration of hydrodynamic models of expanding winds into a surrounding medium with a density gradient, such as those of \cite{wareing07}, it seems likely that the explosion site could be significantly closer to the apex of the bow shock represented by the partial shell (e.g. west of the currently assumed center; see Fig. 3 of \cite{wareing07}).

\section{Summary \label{sec:summary}}

We report new, high spatial resolution \hst/WFC3 imagery of N103B in the LMC, and \hst/COS FUV spectroscopy of the brightest radiative emission knot near the projected center of the SNR.  The images show the extent and structure of the faint Balmer-dominated filaments from the primary blast wave that form a partial shell, as well as bright clumpy radiative shock filaments projected within this shell that represent secondary shocks being driven into density enhancements.  These radiative filaments show a range of relative line intensities that likely arise both from a range of physical conditions (e.g.\ densities and shock velocities) and effective ages (times since encountering the primary blast wave). Some of these filaments represent material in transition to becoming fully radiative, while others appear to have established more complete cooling and recombination zones.  

Our COS data for the brightest radiatively-emitting knot, which itself shows complex structure at \hst\ resolution, presents a rich FUV spectrum and faint underlying continuum.  Modeling the emission-line strengths, along with existing optical data for this filament, shows the emission to be dominated by fast radiative shocks into material with abundances consistent with local LMC ISM.  We point out the similarity between N103B and Kepler's SNR, where the conclusion that the material being encountered by the shock is circumstellar material shed by the precursor system seems secure, and consider this to be likely for N103B as well. Under this assumption, the normal abundances seen imply a precursor system that did not produce an enhancement in abundances at a level detectable in our measurements. We encourage additional effort to search for and/or confirm a possible surviving companion star as a way to better understand the progenitor to this Type Ia supernova.

\acknowledgments

Partial support for the analysis of the data was provided by NASA through grant number HST-GO-14359 from the Space Telescope Science Institute, which is operated by AURA, Inc., under NASA contract NAS 5-26555. CHIANTI is a collaborative project involving George Mason University, the University of Michigan (USA), University of Cambridge (UK) and NASA Goddard Space Flight Center (USA). PFW acknowledges additional support from the NSF through grant AST-1714281.  WPB acknowledges partial support from the JHU Center for Astrophysical Sciences. IRS acknowledges funding from the Australian Research Council under grant FT160100028.


\vspace{5mm}
\facilities{HST(WFC3), HST(COS), ANU 2.3m(WiFeS) }



\software{DS9 \citep{ds9}, IRAF, QFitsView \citep{ott12}}

\pagebreak

\bibliographystyle{aasjournal}

\bibliography{ms}

\begin{center}
\appendix
\section{Chemical Abundance Considerations}
\label{sec:appendix}
\end{center}

In this Appendix, we provide a more detailed discussion of the chemical abundances and their uncertainties.  The models shown in Table 2 assumed the abundance set derived by \cite{dopita19} for the LMC ISM. Here we use the results of the models compared with the extinction corrected line strengths to assess how well these assumed abundances have worked for N103B.

\noindent
{\bf N to O ratio from UV lines:}  The UV lines of N~IV] and N~V and of O~III] and O~IV] give the N:O ratio.  The gas must cool through the temperature range where these lines are formed, and for shocks above 200 $\kms$ the line ratios depend primarily on the relative abundances and the atomic rates.  The excitation rates for these particular lines should be good to 20\%.  There is also an O~V line in the COS spectral range, but it has a very high excitation potential, equivalent to T = $3.3 \times\ 10^5$ K, so the uncertainty is higher for that line.  With the \cite{dopita19} abundances, the N line ratios to O~III] agree with the models, while the ratios to O~IV] are about 20\% too low. 
We estimate that the N:O ratio is  $1.1 \pm\ 0.2$ times the \cite{dopita19} value.

\noindent
{\bf N to H ratio from optical lines:} The \nii\ to \ha\ ratio depends not only on atomic rates, but also on the spectrum of ionizing radiation, on radiative transfer, and on the abundances of the elements that provide competing radiative cooling.  There is also a contribution from the photoionization precursor that was not included in \cite{dopita19} or our analysis.  The magnetic field strength determines the compression ratio, and therefore the ionization parameter in the photoionized region where \nii\ is formed, and this parameter is not independently known.  The predicted ratios of the three models in Table 2 vary from 0.40 to 0.70, compared with an observed value of 0.55.  Considering that the three models do not cover a full range of parameter space, along with reasonable guesses at the uncertainties in photoionization cross sections and radiative transfer, we conclude that the N:H ratio from the optical spectrum is consistent with the \cite{dopita19} value, but it could be as much as 50\% higher.

\noindent
{\bf C to O ratio from UV lines:} The C~II line is likely to be affected by the LMC ISM absorption line, but the C~III] and C~IV lines are not.  The ratios are again determined by the relative abundances and the atomic rates, though the C~III] line can be somewhat dependent on the photoionization.  The ratio of C~IV to O~III] suggests a depletion of 20 to 30\% compared to the D19 abundance ratio, while the ratio to O IV] suggests very little depletion. We conclude that the C:O ratio is lower than the \cite{dopita19} value by a factor of $0.8 \pm\ 0.4$.

\noindent
{\bf Si to O ratio from UV lines:} The Si~III] to O~III] ratio is consistent with \cite{dopita19}, but the Si~III] to O~IV] ratio suggests a 30\% depletion.  The Si~IV to O~IV] ratio suggests a stronger depletion, closer to a factor of 10.  The Si~IV line is blended with the O~IV] multiplet, and it may be subject to a larger uncertainty. We estimate the Si:O ratio to be $0.5 \pm\ 0.3$ times the \cite{dopita19} value.

All of the UV lines used above are moderately strong, but even modest errors get magnified somewhat when looking at ratios.  If we adopt measurement uncertainties of 25\% in the UV line ratios and add in quadrature, we arrive at the following abundance estimates relative to \cite{dopita19} (represented by D19).  N:O =  D19 $\times$ $1.1 \pm\ 0.35$; C:O  =  D19 $\times$ $0.8 \pm\ 0.47$; Si:O =  D19 $\times$ $0.5 \pm\ 0.40$; and from the optical, N:H = D19 $\times$ $1.0 \pm\ 0.5$.


\clearpage

\begin{deluxetable}{rccc}
\tablewidth{0pt}
\tablecaption{HST Observations of N103B}

\tablehead{
\colhead {Instrument} &
\colhead {Filt/Gr} &
\colhead {Dithers} &
\colhead {Total Exp (s)} 
}
\startdata
WFC3-UVIS & F657N & 4   & 2979 \\
WFC3-UVIS & F673N & 4   & 1982 \\
WFC3-UVIS & F502N & 4   & 2051  \\
WFC3-UVIS & F547M & 4   & 782  \\
WFC3-IR   & F164N & 4   & 2012 \\
WFC3-IR   & F160W & 3   & 339  \\
COS       & G140L & (4)\tablenotemark{a} & 2639 \\
\enddata

\tablenotetext{a}{FP-POS positions.}
\label{hst_obsns}
\end{deluxetable}

\begin{deluxetable}{rccccrrrr}
\tablewidth{0pt}
\tablenum{2}
\tablecaption{HST/COS and Key Optical\tablenotemark{a} Emission Line Fluxes for N103B Compared to Models}
\tablehead{
\colhead {Ion} &  
\colhead {$\lambda$\tablenotemark{b}} &  
\colhead {Flux} &  
\colhead {Width\tablenotemark{c}} &  
\colhead {Intensity\tablenotemark{d}} &  
\colhead {Scaled} &
\colhead {Model A\tablenotemark{f}} &
\colhead {Model B\tablenotemark{f}} &
\colhead {Model C\tablenotemark{f}} \\
\colhead {} &  
\colhead {(\AA)} &  
\colhead {($\rm ergs~cm^{-2}~s^{-1}$)} &  
\colhead {(\AA)} &  
\colhead {($\rm ergs~cm^{-2}~s^{-1}$)} &  
\colhead {Int.\tablenotemark{e}} &
\colhead {V=200} &
\colhead {V=220} &
\colhead {V=250} 
}
\startdata
[Ne V]     & 1147.25  & 3.3$\pm$0.6E-16 &  2.2$\pm$0.2  & 4.0E-14 &   14 &    67 &    71 & 74 \\
C III      & 1175.94  & 7.1$\pm$0.7E-16 &  5.1$\pm$0.3  & 6.8E-14 &   25 &    36 &    17 & 19 \\
Si III?    & 1201.18  & 2.5$\pm$0.5E-15 &  8.7$\pm$1.0  & 2.0E-13 &   77 &    88 &    84 & 87 \\
N V        & 1241.82  & 2.3$\pm$0.3E-15 &  5.1$\pm$0.1  & 1.4E-13 &   57 &    69 &    76 & 75 \\
C II       & 1336.69  & 2.5$\pm$0.7E-15 &  4.1$\pm$0.1  & 9.8E-14 &   43 &    94 &   102 & 110 \\
O V        & 1373.20  & 1.6$\pm$0.2E-15 &  4.6$\pm$0.1  & 5.5E-14 &   25 &    35 &    33 & 35 \\
Si IV      & 1395.29  & 3.0$\pm$0.5E-15 &  3.9$\pm$0.1  & 9.6E-14 &   44 &    23 &    17 & 22\\
O IV]      & 1404.40  & 9.6$\pm$1.3E-15 &  6.7$\pm$0.3\tablenotemark{g}  & 3.0E-13 & 137 & 225 & 206 & 221 \\
\,[Si VIII] & 1441.95 & 6.0$\pm$0.8E-16 &  3.4$\pm$0.3  & 1.7E-14 &    8 &   0.4 &     1 & 25 \\
N IV]      & 1484.75  & 1.5$\pm$0.3E-15 &  8.2$\pm$0.5  & 3.8E-14 &   18 &    18 &    17 & 17 \\
C IV       & 1550.59  & 2.7$\pm$0.3E-14 &  5.3$\pm$0.05 & 6.1E-13 &  294 &   402 &   376 & 394 \\
Fe II      & 1603.17  & 1.0$\pm$0.1E-15 &  2.6$\pm$0.1  & 2.1E-14 &   10 &    -  &    -  & - \\
He II      & 1642.03  & 1.9$\pm$0.3E-14 &  4.4$\pm$0.05 & 3.9E-13 &  189 &   175 &   232 & 342 \\
O III]     & 1666.00  & 1.0$\pm$0.1E-14 &  6.5$\pm$0.2\tablenotemark{g}  & 2.0E-13 & 100 & 100 & 100 & 100 \\
Si II      & 1817.39  & 2.6$\pm$0.4E-15 &  4.9$\pm$0.2  & 5.0E-14 &   26 &     3 &     4 &  6 \\
\,[Si III  & 1885.23  & 3.2$\pm$0.7E-15 &  4.1$\pm$0.2  & 6.3E-14 &   31 &    17 &    43 & 39 \\
Si III]    & 1894.41  & 4.6$\pm$0.6E-15 &  5.1$\pm$0.2  & 9.1E-14 &   44 &    35 &    39 & 39 \\
C III]     & 1909.74  & 1.9$\pm$0.3E-14 &  5.1$\pm$0.1  & 3.8E-13 &  188 &   188 &   241 & 268 \\
Cont.\tablenotemark{h}   & (1450) & 3.84E-13 & -  &  1.01E-11 & 4960 &   920 &  1050  &   1730 \\[3pt]
\hline \\[-9pt]
\,[O II]     & 3727       & 8.72$\pm$0.14E-15  & 4.54$\pm$0.05  & 3.71E-14  &   206   &   55 &   189 & 180 \\
\,[Ne III]   & 3869       & 1.53$\pm$0.03E-15  & 3.19$\pm$0.04  & 6.30E-15  &    35   &   70 &    44 &  49 \\
\,[S II]     & 4075       & 2.79$\pm$0.03E-15  & 3.20$\pm$0.05  & 1.09E-14  &    61   &   58 &    33 &  38 \\
He II        & 4686       & 2.83$\pm$0.08E-16  & 3.74$\pm$0.04  & 9.06E-16  &     5   &    8 &    12 &  12 \\
H$\beta$     & 4861       & 6.01$\pm$0.03E-15  & 3.86$\pm$0.03  & 1.80E-14  &   100   &  100 &   100 & 100 \\
\,[O III]    & 5007       & 5.13$\pm$0.04E-15  & 4.11$\pm$0.04  & 1.47E-14  &    82   &  150 &   245 & 222 \\
\,[O I]      & 6300       & 6.79$\pm$0.08E-15  & 4.12$\pm$0.03  & 1.55E-14  &    86   &   65 &    50 &  75 \\
\,H$\alpha$  & 6563       & 2.45$\pm$0.02E-14  & 4.83$\pm$0.06  & 5.31E-14  &   295   &  295 &   295 & 295 \\
\,[N II]     & 6584       & 4.33$\pm$0.06E-15  & 4.49$\pm$0.20  & 9.31E-15  &    52   &   40 &    65 &  70 \\[3pt]
\,[S II]\tablenotemark{i}     & 6725       & 5.45$\pm$0.08E-15  & 4.40$\pm$0.04  & 1.15E-14  &    64   &   54 &    86 &  83 \\
\,[Fe II]\tablenotemark{j} & 16440  & 4.80E-15  &   -    & 5.76E-15  &    32   &  277 &   279 & 346 \\
\enddata
\tablenotetext{a}{Optical line intensities scaled per square arcsec from Dopita et al. (2019) WiFeS spectrum with our adopted extinction correction applied; see text.}
\tablenotetext{b}{Measured central wavelength.}
\tablenotetext{c}{FWHM from Gaussian fit to each line.}
\tablenotetext{d}{Intrinsic flux after correction for extinction; see text.}
\tablenotetext{e}{Relative intensity scaled to I(O~III] 1666)=100 (UV) and I(H$\beta$) = 100 (optical).}
\tablenotetext{f}{For full description of model assumptions, see text.}
\tablenotetext{g}{Blend of multiple components; Gaussian width not representative.}
\tablenotetext{h}{Intensity columns give integrated continuum flux; models give expected hydrogen two-photon continuum flux.}
\tablenotetext{i}{The [S~II] doublet ratio $\lambda$6717/$\lambda$6731 is 0.46, in the high density limit.}
\tablenotetext{j}{[Fe~II] flux from F164N image appropriately scaled to optical spectrum; see text.}
\label{cos_obsns}
\end{deluxetable}

\clearpage


\clearpage


\end{document}